\documentclass[aps,prl,10pt,reprint,superscriptaddress]{revtex4-1}

\usepackage{amsmath,amssymb}
\usepackage[squaren]{SIunits}
\usepackage[usenames,dvipsnames]{color}
\usepackage[pdftex]{graphicx}
\usepackage[normalem]{ulem}
\usepackage{natbib}
\usepackage{hyperref}
\usepackage{bbold}

\definecolor{myGC}{rgb}{0.,0.45,0.25} 
\definecolor{myGCcom}{rgb}{0.,0.8,0.1}

\newcommand{\ave}[1]{\ensuremath{\langle#1\rangle}}

\newcommand{\ket}[1]{\ensuremath{\left|#1\right>}}

\newcommand{\myomit}[1]{}

\newcommand{\tmeas}{t_{\rm e}}
\newcommand{\trel}{t_{\rm r}}

\begin{document}

\title{Entanglement and extreme planar spin squeezing}

\newcommand{\ICFOAddress}{ICFO-Institut de Ciencies Fotoniques, The Barcelona Institute of Science and Technology, 08860 Castelldefels (Barcelona), Spain}
\newcommand{\ICREAAddress}{ICREA -- Instituci\'{o} Catalana de Re{c}erca i Estudis Avan\c{c}ats, 08015 Barcelona, Spain}
\newcommand{\BilbaoAddress}{Department of Theoretical Physics, University of the Basque Country UPV/EHU, P.O. Box 644, E-48080 Bilbao, Spain}

\newcommand{\IKERBASQUEAddress}{IKERBASQUE, Basque Foundation for Science, E-48011 Bilbao, Spain}

\newcommand{\HungaryAddress}{Wigner Research Centre for Physics, Hungarian
Academy of Sciences, P.O. Box 49, H-1525 Budapest, Hungary}

\newcommand{\IQOQIAddress}{Institute for Quantum Optics and Quantum Information (IQOQI), Austrian Academy of Sciences, Boltzmanngasse 3, A-1090 Vienna, Austria}

\author{G.~Vitagliano}
\email[]{giuseppe.vitagliano@univie.ac.at}
\affiliation{\IQOQIAddress}
\affiliation{\BilbaoAddress}

\author{G.~Colangelo}
\email[]{giorgio.colangelo@alumni.icfo.eu}
\affiliation{\ICFOAddress}

\author{F.~Martin Ciurana}
\affiliation{\ICFOAddress}

\author{M.~W.~Mitchell}
\affiliation{\ICFOAddress}
\affiliation{\ICREAAddress}

\author{R.~J. Sewell}
\affiliation{\ICFOAddress}

\author{G.~T\'oth}
\affiliation{\BilbaoAddress}
\affiliation{\IKERBASQUEAddress}
\affiliation{\HungaryAddress}

\date{\today}

\newcommand{\omegaL}{\omega_{\rm L}}
\newcommand{\qed}{\ensuremath{\hfill \Box}}

\newcommand{\J}{\ensuremath{\mathbf{J}}}

\begin{abstract}
We introduce an entanglement-depth criterion optimized for planar quantum squeezed (PQS) states. It is connected with the sensitivity of such states for estimating a phase generated by rotations about an axis orthogonal to its polarization. 
We compare numerically our criterion with the well-known extreme spin squeezing condition of 
S\o rensen and M\o lmer [\href{http://link.aps.org/doi/10.1103/PhysRevLett.86.4431}{Phys. Rev. Lett. 86, 4431 (2001)}] 
and show that our condition detects a higher depth of entanglement when both planar spin variances are squeezed below the standard quantum limit. 
We employ our theory to monitor the entanglement dynamics in a PQS state produced via quantum non-demolition (QND) measurements using data from a recent experiment [\href{https://doi.org/10.1103/PhysRevLett.118.233603}{Phys. Rev. Lett. 118, 233603 (2017)}].
\end{abstract}

\maketitle
{\it Introduction.---}Detecting entanglement in large quantum systems is a major goal in quantum information science and underpins the development of quantum technologies~\cite{horodeckirev,Guhne2009Entanglement}. 
Attention has now shifted toward the practical use of entanglement as a resource: in 
particular, entanglement-enhanced sensing using ensembles of $10^3-10^{12}$ atomic spins has emerged as a major application~\cite{tothapellaniz14,2016arXiv160901609P}.
In this context, spin-squeezing inequalities can be used to quantify entanglement-enhanced sensitivity.
Standard treatment studies spin-squeezed states (SSS),
characterized by a large spin polarization in the $y$-direction and a small variance in the $z$-direction, via the parameter $\xi_s^2:=~\tfrac{N(\Delta J_z)^2}{|\ave{J_y}|^2}$, where $J_v=\sum_{n=1}^N j_v^{(n)}$ for $v=x,y,z$ are the collective spin components,  
$j_v^{(n)}$ are single particle spin operators, 
$N$ is the total number of atoms. 
Let us assume that the mean collective spin points in the $y$-direction, while $J_z$ is the spin component in the orthogonal direction with the smallest variance.
Then, states with $\xi_s^2 < 1$ provide quantum enhanced sensitivity 
for estimating phases $\phi \approx 0$ due to small rotations around $J_x$~\cite{Kitagawa1993Squeezed,Wineland1994Squeezed}.
Such states have been produced using various platforms, including cold atoms~\cite{Hald1999Spin,Fernholz2008Spin,Orzel2386,Riedel2010Atom-chip-based,Esteve2008Squeezing,SchleierSmithPRL2010,Leroux2010,Gross2010Nonlinear,BohnetJ2014,SewellPRL2012,Cox2016,HostenN2016}, trapped ions~\cite{Meyer2001}, magnetic systems~\cite{Auccaise2015} and photons~\cite{MitchellExtremespinsqueezing2014}.

The metrological sensitivity is strongly connected to entanglement.
According to the original spin squeezing criterion, $\xi_s^2 < 1$ also implies entanglement for atoms with spin $j=1/2$~\cite{Sorensen2001Many-particle}. 
This was extended to systems with a higher spin by S\o rensen and M\o lmer, who introduced a method to quantify entanglement by means of the so-called {\it depth of entanglement}, i.e., the number of particles in the largest separable subset \cite{Sorensen2001Entanglement}. 
Several other highly entangled states have recently been found useful for quantum metrology, such as Dicke states. These states are unpolarized, have a large value of $\ave{J_x^2+J_y^2}$, and a small variance in the $z$-direction. 
Dicke states can be included in a class of generalized spin squeezed states, and their degree of entanglement can be quantified through collective spin variances. 
Spin squeezing inequalities have been developed to characterize entanglement in such states~\cite{Lucke2014Detecting,vitagliano16}, which have been produced in experiments with photons~\cite{Wieczorek2009Experimental,Prevedel2009Experimental} and Bose-Einstein condensates~\cite{Lucke2011Twin,Hamley2012Spin-nematic,Lucke2014Detecting,Hoang2016Characterizing,Luo620}.

Here, we focus on a different class of states, called {\it planar quantum squeezed} (PQS) states, studied theoretically in~\cite{HePRA2011,He2012,PuentesNJP2013}, and observed in a recent experiment~\cite{ColangeloNat2017,ColangeloPQS2017}.
These states have reduced spin variances in two directions, i.e.,
$(\Delta J_{\parallel})^2:=(\Delta J_y)^2+(\Delta J_z)^2$ is small,
and a large in-plane polarization, i.e., $\ave{J_y}\approx Nj.$ They
provide quantum-enhanced sensitivity in estimating phases generated by rotations about the $\hat x$ axis, without the need of first localizing the phase around $\phi\simeq0$ and are useful for tasks such as tracking a changing phase shift, or simultaneous estimation of phase and amplitude beyond classical limits \cite{ColangeloNat2017,ColangeloPQS2017}.
The planar squeezing parameter 
\begin{equation}\label{eq:xi_parallel}
\xi^2_{\parallel}:=~\frac{(\Delta J_{\parallel})^2}{|\ave{J_\parallel}|},
\end{equation} 
where $|\ave{J_\parallel}|:=\sqrt{\ave{J_y}^2+\ave{J_z}^2}$ is the in-plane polarization, was introduced by He \textit{et al.}~\cite{HePRA2011,He2012} to quantify such enhanced sensitivity and detect entanglement. 
While planar squeezed states have intriguing properties, the relation between their metrological usefulness
and their multiparticle entanglement has not been explored so far.

In this paper, we show how to detect multipartite entanglement that is also useful metrologically for the phase estimation task above.
In particular, we introduce a method to detect the depth of entanglement based on the planar squeezing parameter $\xi_{\parallel}^2.$ We present the condition
\begin{equation}\label{eq:plansqcritk}
\xi^2_{\parallel} \geq \zeta^2_J,
\end{equation}
where $\zeta^2_J$
is the minimum value of the planar squeezing parameter {\it over single particle states} of spin $J.$ We prove that
for all spin-$j$ ensembles that contain groups of at most $k$-entangled particles, called $k$-producible~\cite{Guhne2005Multipartite,Sorensen2001Entanglement}, Eq.~\eqref{eq:plansqcritk} holds with $J=kj.$
Thus, $\xi^2_{\parallel}< \zeta^2_J$, implies a depth of entanglement of at least $k+1=J/j+1.$ We can even estimate at least how many particles must be in fully entangled $(k+1)$-particle groups. 
We stress that our criterion is very simple to use. 
We need to calculate $\zeta^2_J$ only once for the relevant range of $J,$ then Eq.~\eqref{eq:plansqcritk} can be applied for entanglement detection without any additional numerical optimization.

\begin{figure}
\includegraphics[width=\columnwidth ]{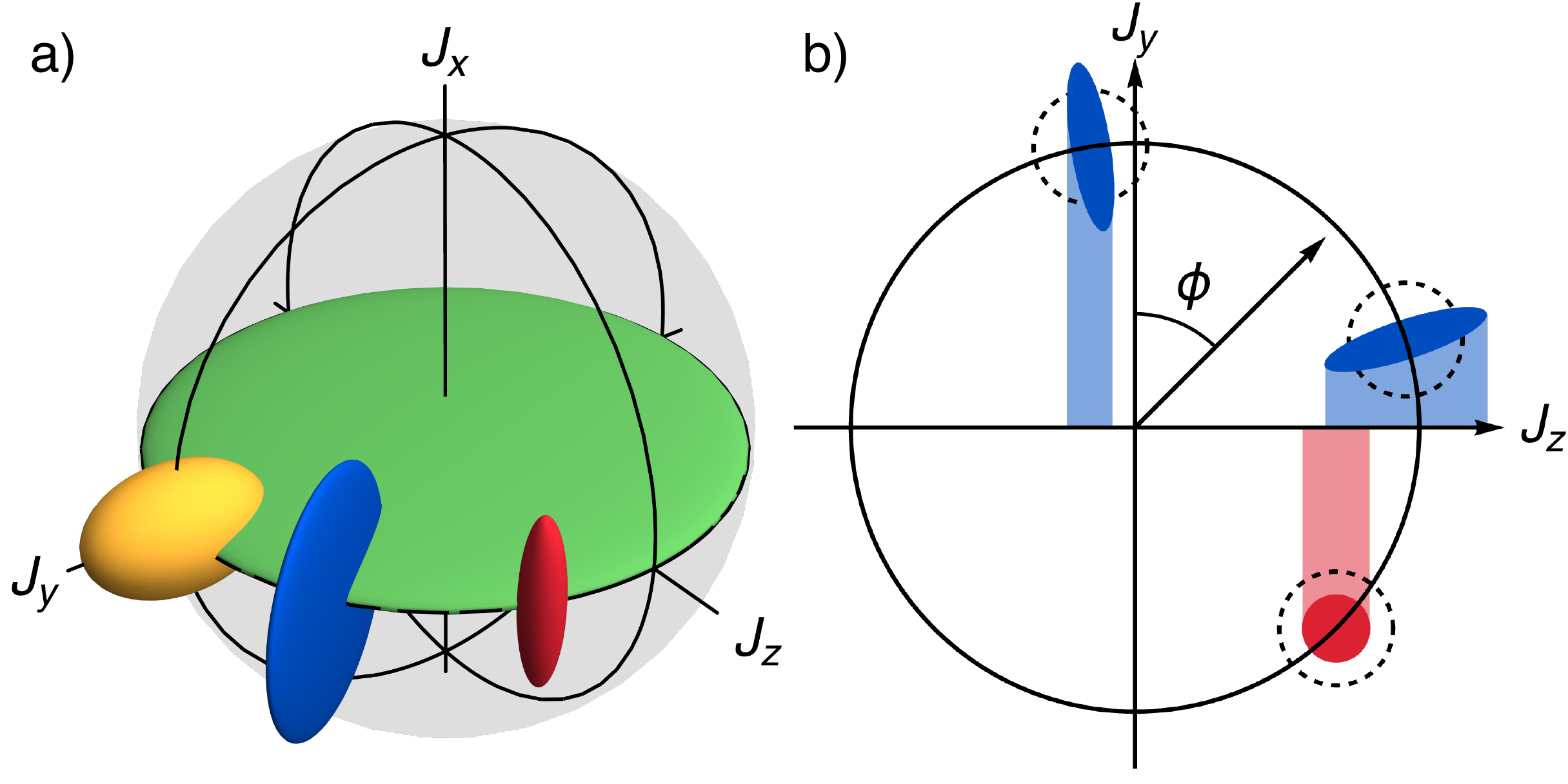}
\caption{
a) Bloch-sphere representation of:
(yellow) spin-coherent states,  
including uncertainty in the length of the spin vector arising from Poissonian fluctuations in preparing atomic ensemble;
(blue) SSS produced by squeezing the $J_z$ variance of the spin-coherent state;
(red) PQS state produced by squeezing both $J_z$ and $J_y$;
(green) a Dicke state.
b) Sensitivity advantage of PQS state compared to SSS in detecting an unknown phase $\phi$.
The dashed black circles indicate the classical limit for independent spins $\Delta\phi=1/\sqrt{N}$.
The SSS provides enhanced sensitivity for detecting phases around $\phi\simeq0$, but reduced sensitivity for phases around $\phi\simeq\pm\pi/2$.
In contrast, while the sensitivity of the PQS state is slightly worse around $\phi\simeq0$, it provides enhanced sensitivity for all phases $0\le\phi\le2\pi$.
\label{fig:state}}
\end{figure}

Finally, we examine the usefulness of our criterion. We compare it
to the S\o rensen-M\o lmer criterion numerically, and find that it detects a higher entanglement depth for a large class of quantum states. We also test our theory using data from a recent experiment in which a PQS state was generated via semi-continuous quantum non-demolition (QND) measurement~\cite{ColangeloNat2017,ColangeloPQS2017}.

{\it Link between our parameter and metrological sensitivity.---}We consider a protocol in which a collective spin state is rotated about the $J_x$ axis and accumulates a phase $\phi$ such that the operator $J_z$ evolves as $J_z^{\rm out} = J_z^{\rm in} \cos \phi - J_y^{\rm in} \sin \phi$. Afterwards, the phase is inferred from repeated measurements of $J_z^{\rm out}$, with a sensitivity given by the error-propagation formula $(\Delta \phi)^2 = (\Delta J_z^{\rm out})^2/|\partial_\phi \ave{J_z^{\rm out}}|^2$. 
We consider as a reference an input state with an uncertainty at the standard quantum limit (SQL), $(\Delta J_y^{\rm in})_{\rm SQL}^2=(\Delta J_z^{\rm in})_{\rm SQL}^2=\tfrac 1 2 |\ave{J_\parallel^{\rm in}}|$~\cite{PuentesNJP2013,ColangeloPQS2017}.
A SQL-limited state cannot beat the shot-noise limit corresponding to separable states, since $(\Delta \phi)_{\rm SQL}^2=|\ave{J_\parallel^{\rm in}}|/(\ave{J_z^{\rm in}}^2\cos^2 \phi + \ave{J_y^{\rm in}}^2 \sin^2 \phi) \ge  1/N$. Hence, we can normalize the sensitivity with respect to the SQL and obtain that
$(\Delta \phi)^2/(\Delta \phi)_{\rm SQL}^2=\left[(\Delta J_z^{\rm in})^2\cos^2 \phi+ (\Delta J_y^{\rm in})^2\sin^2 \phi\right]/|\ave{J_\parallel^{\rm in}}|$.
By averaging over the phase we find
\begin{equation}\label{eq:avsensitivity}
\int_0^{2\pi} \tfrac{\rm d \phi}{2\pi} (\Delta \phi)^2/(\Delta \phi)_{\rm SQL}^2 = \tfrac 1 2 \xi^2_\parallel.
\end{equation}
Thus, the planar squeezing parameter quantifies the average sensitivity enhancement compared to the SQL over the interval $0\le\phi\le2\pi$.

{\it Entanglement criterion for planar squeezing.---}Following an approach similar to past works~\cite{Sorensen2001Entanglement,vitagliano16}, we derive a tight criterion to detect the depth of entanglement based on the above planar squeezing parameter by computing the function
\begin{equation}\label{eq:hkjfunc}
G^{(j)}_{k}(X):= \tfrac 1 {kj} \min_{\stackrel{\phi \in (\mathbb C^d)^{\otimes k}}{\tfrac{1}{kj} \ave{L_y}_\phi = X}}  \left[ (\Delta L_y)_\phi^2+(\Delta L_z)_\phi^2 \right] ,
\end{equation}
where $d=2j+1$, $j$ is the single-particle spin quantum number,
$L_v$ are {\it $k$-particle operators} constructed from spin-$j$ operators, i.e., $L_v = \sum_{n=1}^k j_v^{(n)},$ where $j_v^{(n)}$ are single-particle spin components.
First, we derive a criterion that contains a tight lower bound on the planar spin variance valid for all states with a depth of entanglement smaller than $k$.

{\bf Observation 1.---}Every $k$-producible state of a spin-$j$ particle system with average number of particles $\ave{N}$ must satisfy the tight inequality
\begin{equation}\label{eq:plansqcritkprovv}
(\Delta J_{\parallel})^2 \geq \ave{N}j \mathcal G^{(j)}_k\left( \tfrac{|\ave{J_\parallel}|}{\ave{N}j} \right) ,
\end{equation}
where $\mathcal G^{(j)}_k$ is defined as the convex hull of (\ref{eq:hkjfunc}).
Thus, every state that violates Eq.~(\ref{eq:plansqcritkprovv}) must have a depth of entanglement of at least $k+1.$ 

{\it Proof.---}For pure $k$-producible states of a fixed number  of particles $N$ we have 
$(\Delta J_\parallel)_N^2= \sum_{n} \left[ (\Delta L^{(n)}_y)^2 + (\Delta L^{(n)}_z)^2 \right] \geq \sum_n k_nj 
\mathcal G^{(j)}_{k_n}( \ave{L_y^{(n)}}/k_nj)$,
where the $L_v^{(n)}$ are collective operators of $0\leq k_n \leq k$ particles. The second inequality follows directly from the definition
of $\mathcal G^{(j)}_{k_n}$. Now, we have to take into account the fact that $\mathcal G^{(j)}_{k_n}$ are (i) convex and (ii) decreasing for increasing the index, i.e., $\mathcal G^{(j)}_{r}\leq \mathcal G^{(j)}_s$ for $r\geq s$ and that $k\geq k_n$ and $\sum_n k_n=N.$ Then, follows
$\sum_n k_nj  \mathcal G^{(j)}_{k_n }(\ave{L_y^{(n)}}/k_nj ) \geq \sum_n k_nj  \mathcal G^{(j)}_k(\ave{L_y^{(n)}}/k_nj )
\geq Nj  \mathcal G^{(j)}_k( \ave{J_y}/Nj),$
where the first inequality comes from property (ii) and the second from property (i) and Jensen's inequality. 
Clearly, if $N$ is divisible by $k$ then the inequality \eqref{eq:plansqcritkprovv} is tight by construction.

Let us consider now a state with a non-zero particle number variance given  as
$\varrho=\sum_N Q_N \varrho_N$, where $\varrho_N$ are states with a fixed particle number $N$ and $Q_N$ are the corresponding probabilities.
From the properties above and from the concavity of the variance follows also that for such a state $(\Delta J_\parallel)^2\geq \sum_N Q_N (\Delta J_\parallel)_N^2 \geq \sum_N Q_N Nj  \mathcal G^{(j)}_k( \ave{J_y}/Nj) \geq \ave{N}j \mathcal G^{(j)}_k(\ave{J_y}/\ave{N}j )$ holds, $\ave{N}=\sum_N Q_N N$ being~the~average~particle~number. \qed

Starting from the function $G^{(j)}_{k}(X)$ given in Eq.(\ref{eq:hkjfunc}) we can construct numerically the convex hull $\mathcal G_k^{(j)}$ that will have properties (i) and (ii) mentioned in the proof, as we now explain.

\begin{figure}[ht!]
\includegraphics[width=0.9\columnwidth]{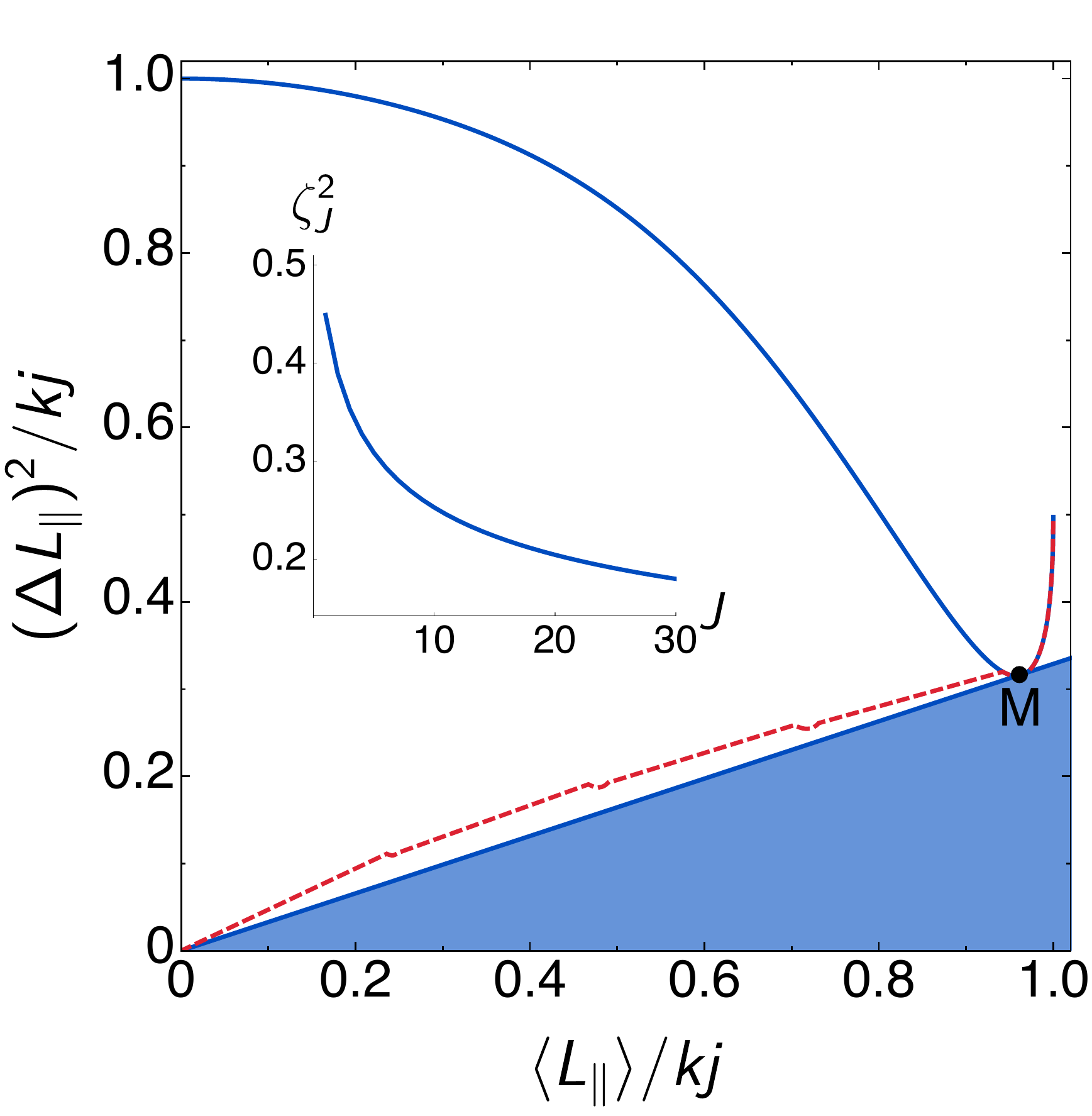}
\caption{(color online). 
Lower bounds to $(\Delta L_\parallel)^2/kj$ as functions of $\ave{L_\parallel}/kj$ for a system of $k$ spin-$j$ particles. The case of $k=4$ and $j=1$ is shown. 
(dashed) The function $G_{4}^{(1)}(X)$.
(solid) The function $G_{4}^{\rm sy}(X)$ computed on the symmetric subspace.
The convex hull of $G_{4}^{(1)}(X),$ denoted by $\mathcal G_{4}^{(1)}(X)$ is a linear function for $X\leq X_{{\rm min}}={\rm argmin} [G^{\rm sy}_4(X)/X],$ while for $X>X_{{\rm min}}$ it coincides with $G^{\rm sy}_4(X).$
$M$ denotes the point of the curve for which $X=X_{{\rm min}}.$
The straight line provides a lower bound on $\mathcal G_{4}^{(1)}(X)$. 
(inset) The parameter $\zeta^2_J$ as a function of $J$. 
\label{fig:gkjspin1plot}}
\end{figure}

{\it Numerical computation of $\mathcal G_k^{(j)}$.---}In order to detect the depth of entanglement with our criterion we need to carry out the optimization in Eq.~(\ref{eq:hkjfunc}). For for $k=1$ and $j=1,$ straightforward algebra yields $G_{1}^{(1)}(X)=\tfrac 3 2 - X^2 - \tfrac 1 2 \sqrt{1-X^2}$. Analytical expressions are very hard to obtain even for the next simplest cases. 
Numerically the optimization can be carried out by finding the ground states of $H_\lambda= L_y^2+L_z^2-\lambda L_y$,
where $L_v$ are the $k$-particle operators defined before, which can be written in the usual block-diagonal form, where each block acts in some spin-$J$ subspace. Hence, the higher the $k$, the more are the spin-$J$ subspaces involved. One can show that, due to these, $\mathcal G_k^{(j)}$ cannot increase with $k$, which is just property (ii) needed in
the proof of Observation 1 (see Supplemental Material \cite{Note1} for more details).  When the optimization in Eq.~(\ref{eq:hkjfunc}) is restricted to the subspace with the maximal spin $J=kj,$ we call the resulting function $G^{\rm sy}_J(X)$, since it is computed by an optimization restricted to the symmetric subspace of $k$-partite states.  In Fig.~\ref{fig:gkjspin1plot},  we present a concrete example. The function $G_{4}^{(1)}(X)$ is plotted together with its convex hull $\mathcal G_{4}^{(1)}(X)$ and $G^{\rm sy}_4(X).$
We can see that a simple linear function corresponding to a straight line 
on the figure 
can be used as a lower bound to $\mathcal G_{4}^{(1)}(X).$ This lower bound works in general, as we show in what follows.

{\it Linear lower bound.---}As outlined above, the computation of $\mathcal G_{k}^{(j)}(X)$ still requires some numerics, which can be hard for high $k$ and $j$. Here, we will simplify further this task by finding a suitable lower bound that requires only the numerical computation of $G^{\rm sy}_{J}(X)$ with $J=kj$ and is thus easier than computing the full $\mathcal G_{k}^{(j)}(X)$. 

{\bf Observation 2.---}A convex lower bound to the curve $G_k^{(j)}(X)$ defined as in Eq.~(\ref{eq:hkjfunc}) is given by
\begin{equation}\label{eq:lowboundJ}
\mathcal G^{(j)}_k(X) \geq X \zeta^2_J ,
\end{equation}
where $\zeta^2_J:=\min_{|\psi_k\rangle} [ (\Delta L_y)_{\psi_k}^2 + (\Delta L_z)_{\psi_k}^2]/\ave{L_y}_{\psi_k}$ is the minimum value of the planar squeezing parameter over single particle states $|\psi_k\rangle$ of spin $J=kj$. The proof is given in the Appendix.

With this method we need only to determine $\zeta^2_J$ for the relevant range of $J$, which can be written as $\zeta^2_J=\min_X [G^{\rm sy}_{J}(X)/X]$. Thus, as a simple algorithm 
one can: i) Find the ground states $|\phi_\lambda\rangle$ of $H_\lambda$ {\it restricted to the symmetric subspace}; ii) compute $(\Delta L_\parallel)_{\phi_\lambda}^2$ and $\ave{L_y}_{\phi_\lambda}$; and finally take $\zeta^2_J=\min_{\phi_\lambda} (\Delta L_\parallel)_{\phi_\lambda}^2/\ave{L_y}_{\phi_\lambda}$, which is feasible until very large $J,$ up to the thousands \cite{Note1}. 
As an example the values of $\zeta^2_J$ up to $J=27$ are given in Table~\ref{tab:ent}, while
the qualitative behavior can be observed in the inset of Fig.~\ref{fig:gkjspin1plot}. Note that Eq.~(\ref{eq:lowboundJ}) is a tight approximation only for $k\geq 2$, independently on $j$. For $k=1$ the original criterion given in Eq.~(\ref{eq:plansqcritkprovv}), has to be used instead.

\setlength{\tabcolsep}{5pt}
\begin{table}[hb]
\caption{Values of $\zeta^2_J$ for $0\leq J\leq 27$.}
\begin{center}
\begin{tabular}{c|c|c|c|c|c}
$J$ & $\zeta_J^2$ & $J$ & $\zeta_J^2$ & $J$ & $\zeta_J^2$ \\
\hline
 1 & 0.45 & 10 & 0.26067 & 19 & 0.21111 \\
2 & 0.44906 & 11 & 0.25262 &20 & 0.20758 \\
3 & 0.38945 & 12 & 0.2455 & 21 & 0.20428 \\
4 & 0.35321 & 13 & 0.23913 & 22 & 0.20118 \\
5 & 0.32779 & 14 & 0.23338 & 23 & 0.19826 \\
6 & 0.30852 & 15 & 0.22815 & 24 & 0.19551    \\
7 & 0.29318 & 16 & 0.22336 & 25 & 0.1929  \\
8 & 0.28054  & 17 & 0.21896 & 26 & 0.19043 \\
9 & 0.26986  & 18 & 0.21489 & 27 & 0.18809 \\
\end{tabular}
\end{center}
\label{tab:ent}
\end{table}%

From Observations~1 and 2,
we immediately obtain Eq.~(\ref{eq:plansqcritk}), which connects  the metrological performance of the  PQS states
to their entanglement depth.
Next, we show that apart from proving that the entanglement depth is $k+1$, 
we can also obtain information about how many particles are in fully entangled groups of $(k+1).$
This provides a simple interpretation of the degree of the violation of Eq.~(\ref{eq:plansqcritk}). 

{\bf Observation 3.---}Let us assume that the total polarization is equally distributed over all particles. Then, there is at least a fraction 
$f_{k+1}=(1-\xi^2_{\parallel}/\zeta^2_J)$
of particles in fully entangled groups of $(k+1)$ or more, with $k$ given by $J/j$. The proof is given in the Supplemental Material, where the case of varying particle numbers is included in the model \cite{Note1}. We discuss that without the assumption of equally split polarization, the above statement still holds for almost totally polarized states, i.e., with $\ave{J_y}\approx Nj$ and that similar ideas work also for the S\o rensen-M\o lmer
criterion \footnote{\label{footnote_supplement}See Supplemental Material at [URL].}.

\begin{figure}[ht!]
\includegraphics[width=\columnwidth]{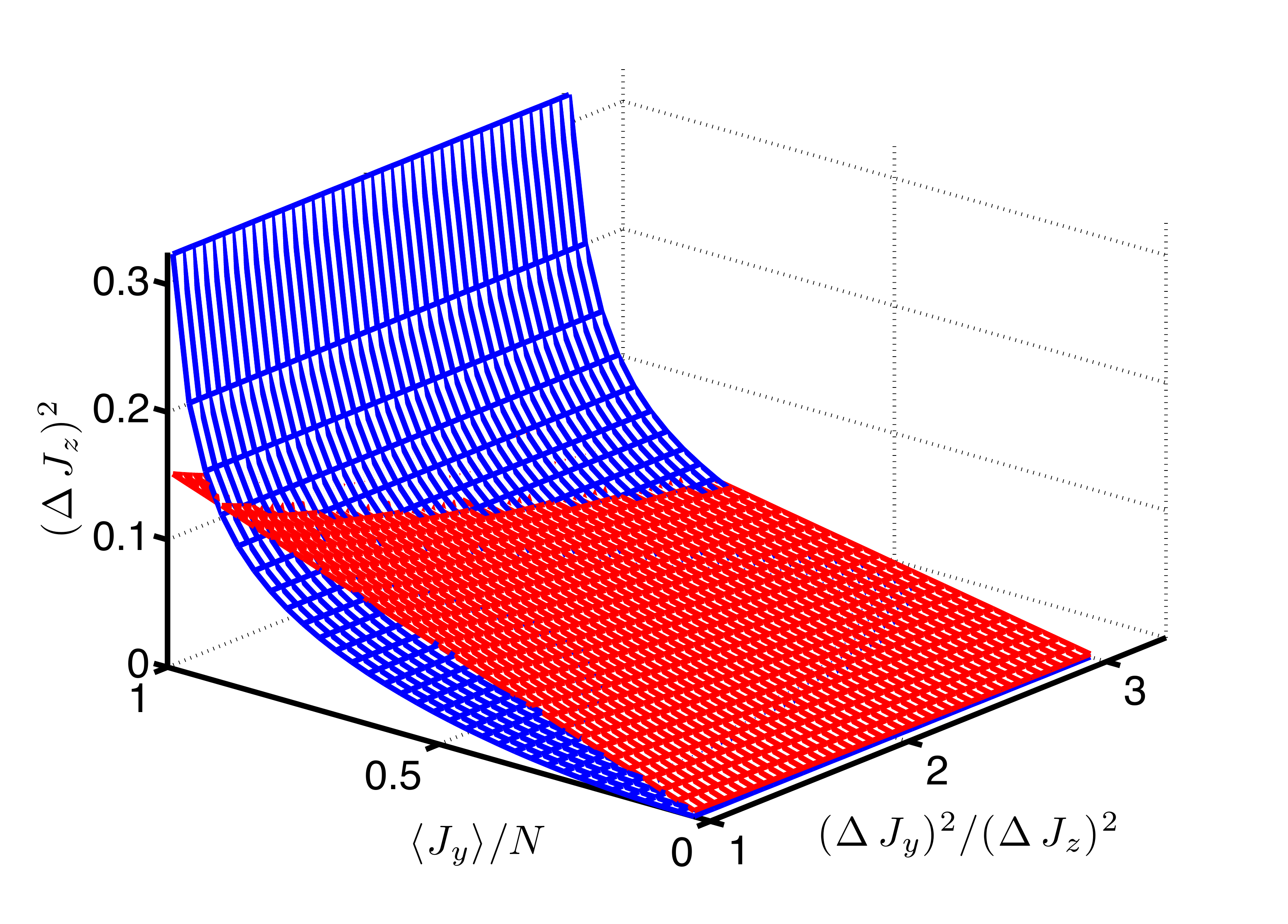}
\caption{(color online) 
Lower bound on $(\Delta J_z)^2$ given by the
(blue) extreme spin squeezing and (red) planar squeezing criteria taken for $k=5$ and $j=1$ as a function of the ratio between the two planar spin variances and the in-plane polarization. Our criterion detects a depth of entanglement higher than S\o rensen-M\o lmer's for the parameter values for which the red plot is above the blue one.
\label{fig:figcomparison}}
\end{figure}

{\it Practical use of the criterion.---}Thus, our criterion can be employed to detect the depth of entanglement in states for which two collective spin variances are known, as well as the total in-plane polarization. 
With the same input information, it would be possible to consider also the S\o rensen-M\o lmer extreme spin squeezing condition \footnote{Note that the S\o rensen-M\o lmer condition requires as input just one variance, that must be the one orthogonal to the polarization.}. Then, we can numerically compare the two criteria and study in which cases our criterion is more suitable to detect entanglement.  
To do this we parametrize the states with the ratio $\alpha=(\Delta J_z)^2/(\Delta J_y)^2$ between the two spin variances, and the total in-plane polarization $\beta=\ave{J_y}/N$. Then, we plot the lower bound on $(\Delta J_z)^2$
for $k=5$ and $j=1$ for various values of $\alpha$ and $\beta$ and see for which regions of the $(\alpha, \beta)$ plane our criterion detects a higher depth. The result is shown in Fig.~\ref{fig:figcomparison}, where we can observe
that our criterion detects a higher degree of entanglement in most of the plane, especially whenever the two variances become equal. Vice versa, totally polarized states that are spin squeezed only along the $z$-direction are detected with a higher depth by S\o rensen-M\o lmer's criterion. All these statements are valid also for other $k$ and $j$ values. Thus, our criterion is especially tailored for detecting 
PQS states and distinguish those from traditional spin squeezed states, which are optimally detected by S\o rensen-M\o lmer's criterion. Furthermore, the linearity of our criterion makes it directly connected with improved sensitivity in phase estimations: a value of $\xi^2_{\parallel}$ below the threshold given by the constant $\zeta^2_J$ with $J=kj$ implies that: (1) the state must be $(k+1)$-entangled and (2) its average sensitivity to rotations about the axis orthogonal to the plane of squeezing as compared to the SQL (given by the parameter $\xi^2_{\parallel}$ itself as in Eq.~(\ref{eq:avsensitivity})) is better than that of any state with depth of entanglement $k$ or lower.

\begin{figure}[ht!]
\includegraphics[width=0.9\columnwidth]{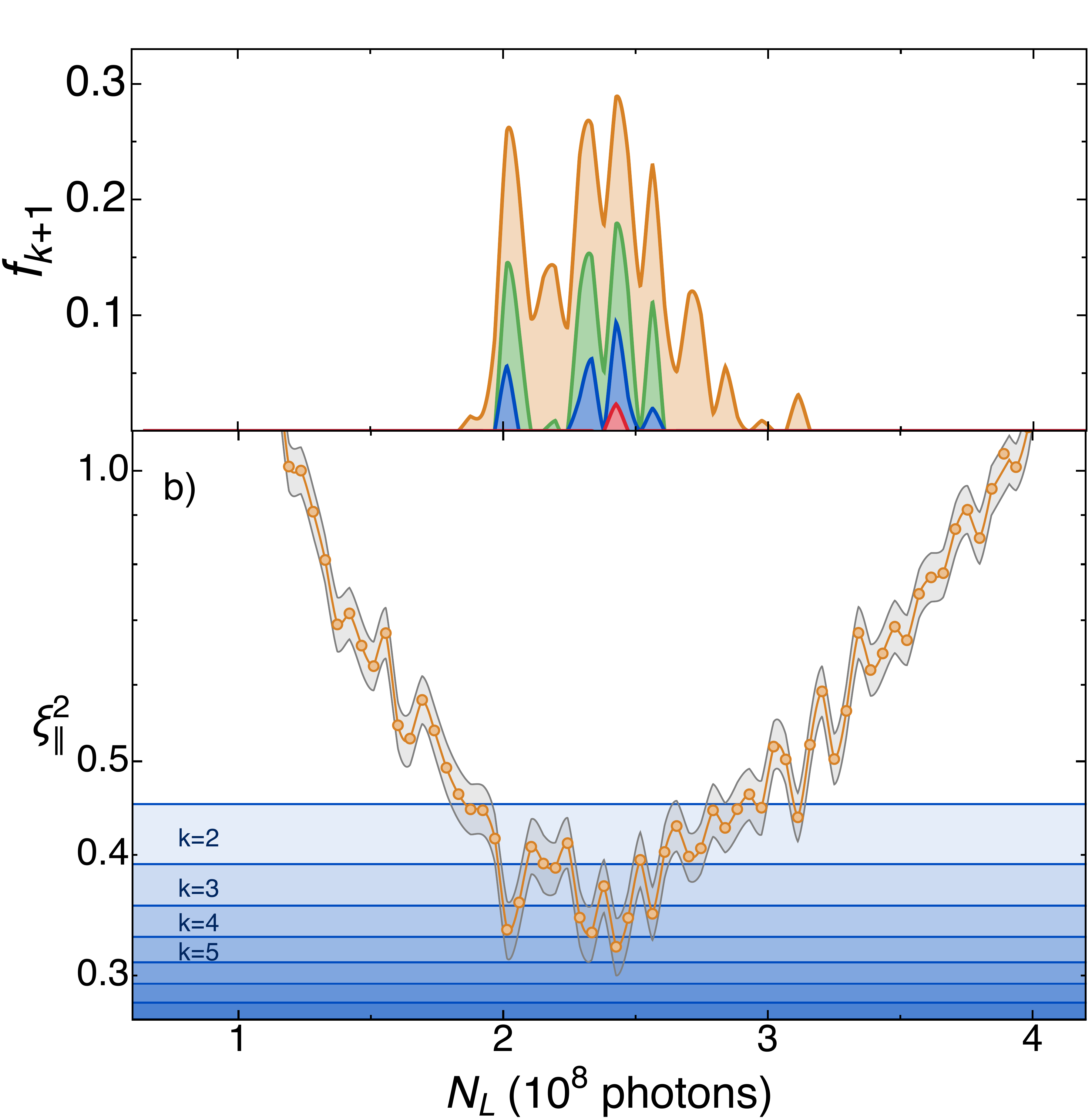}
\caption{Top: Shaded curves, lower bound for number of atoms in fully entangled groups of at least $(k+1)$ particles.
From top to bottom: $k=1$, $2$, $3$ and $4$.
Bottom: Squeezing parameter $\xi_{\parallel}^2$ as function of the number of photons $N_{\rm L}$ used in the QND measurement. 
Orange shaded area represents $\pm 1\sigma$ confidence interval.
\label{fig:fig2}}
\end{figure}

Next, we employ our criterion (\ref{eq:plansqcritk}) to analyze entanglement generated in a recent experiment in which a PQS state was produced in an ensemble of $N=1.75 \times 10^6$ cold $^{87}$Rb atoms via semi-continuous quantum non-demolition (QND) measurements~\cite{ColangeloPQS2017}.
In Fig.~\ref{fig:fig2} we plot the observed planar squeezing parameter $\xi^2_\parallel$ as a function of the measurement strength, parametrized by number of photons $N_{\rm L}$ used in the QND measurement.
As $N_{\rm L}$ increases, the input spin-coherent state evolves into a planar squeezed state, with squeezing observed between $N_{\rm L}\simeq 2 \cdot 10^8$ and $N_{\rm L}\simeq3 \cdot 10^8$ photons, after which the spin variances increase due to noise and decoherence introduced by off-resonant scattering of probe photons.
We also plot the corresponding fraction $f_{k+1}$ of atoms in fully entangled groups of $(k+1)$ or more, detected using our criterion.
We observe the corresponding increase in entanglement depth with $N_{\rm L}$ up to the optimum of $N_{\rm L}~=~2.47 \times 10^8$ photons, after which entanglement is gradually lost.
At the optimum $N_{\rm L}$ we observe a spin coherence $\ave{J_\parallel}~=~0.83~N$, and a planar squeezing parameter $\xi_\parallel^2 = 0.32\pm0.02$.
For comparison, using the criterion developed by He \textit{et al.}~\cite{HePRA2011,He2012}, one would detect a fraction $0.39$ of atoms in entangled states, without any information about the depth of entanglement. The details of the experiment are given in the Supplemental Material \cite{Note1}.

{\it Conclusions.---}We have introduced a new criterion suitable to detect the depth of entanglement in planar squeezed states, and to distinguish them from traditional spin squeezed states, detectable with S\o rensen-M\o lmer criterion~\cite{Sorensen2001Entanglement}. Our criterion is simple to evaluate and directly connected with the sensitivity of PQS states for phase estimations that do not require any prior knowledge of the phase. By numerical comparison, we have also shown that our criterion represents an important alternative to S\o rensen-M\o lmer's suitable to detect entanglement in PQS states. Finally, we tested our criterion with data from a recent experiment in which a PQS state was generated via semi-continuous QND measurement~\cite{ColangeloPQS2017}. 

This work was supported by the EU (ERC Starting Grant 258647/GEDENTQOPT, CHIST-ERA QUASAR, COST Action CA15220, ERC project AQUMET (grant agreement 280169), QUIC (grant agreement 641122) and ERIDIAN (Grant Agreement No. 713682)), the Austrian Science Fund (FWF) through the START project Y879-N27,
the Spanish MINECO/FEDER, MINECO (projects No. FIS2012-36673-C03-03 and No. FIS2015-67161-P, MAQRO (Ref. FIS2015-68039-P), XPLICA (FIS2014-62181-EXP) and Severo Ochoa Programme for Centres of Excellence in R\&D (SEV-2015-0522)),
the Basque Government (Project No. IT4720-10),
the National Research Fund of Hungary OTKA (Contract No. K83858), 
the CERCA Programme / Generalitat de Catalunya, Catalan 2014-SGR-1295, and by Fundaci\'{o} Privada CELLEX.

\appendix

{\it Appendix. Proof of Observation 2.---}Let us consider a general pure $k$-particle state $\ket{\phi}=\sum_J a_J \ket{\psi_J},$
where each $|\psi_J\rangle$ is a single spin-$J$ particle state.
We assume that the mean planar spin points into the $y$-direction.
We now need that the collective $k$-particle spin components $L_v$ can be written as the direct sum of operators $L_v^{(J)}$ acting on spin-$J$ particle spaces with $0\leq J \leq kj$ (or $1/2\leq J \leq kj$ for odd $k$ and half-integer $j$).
Then, since the collective angular momentum operators do not couple the different spin-$J$ subspaces of the total $k$-particle space to each other, we have $(\Delta L_\parallel)_\phi^2 \geq \sum_{J>0} a_J^2 \ave{L_y^{(J)}}_{\psi_J}\tfrac{(\Delta L_\parallel^{(J)})_{\psi_J}^2 }{\ave{L_y^{(J)}}_{\psi_J}}.$  Hence,  $(\Delta L_\parallel)_\phi^2 \geq \sum_{J>0} a_J^2 \ave{L_y^{(J)}}_{\psi_J} \zeta^2_J$ follows.
Finally, we obtain $\sum_{J>0} a_J^2 \ave{L_y^{(J)}}_{\psi_J} \zeta^2_J \geq \zeta^2_{J_{\rm max}} \sum_{J>0} a_J^2 \ave{L_y^{(J)}}_{\psi_J}= J_{\rm max} X \zeta^2_{J_{\rm max}}$, where $J_{\rm max}=kj$, since $X=\tfrac 1 {kj} \ave{L_y}_{\phi}=\tfrac 1 {kj} \sum_{J>0} a_J^2 \ave{L_y^{(J)}}_{\psi_J}$ and
$\zeta^2_J \geq \zeta^2_{J^\prime}$ for $J \leq J^\prime$. The last property can be observed numerically, cf. Fig.~\ref{fig:gkjspin1plot}(inset). Due to the concavity of the variance, the statement follows for mixed states.~\qed 

\bibliographystyle{../biblio/Science}
\bibliography{../biblio/biblio}

\begin{thebibliography}{10}

\bibitem{horodeckirev}
R.~Horodecki, P.~Horodecki, M.~Horodecki, K.~Horodecki, Quantum entanglement,
  {\it Rev. Mod. Phys.\/} {\bf 81}, 865 (2009).

\bibitem{Guhne2009Entanglement}
O.~G{\"u}hne, G.~T{\'o}th, Entanglement detection, {\it Phys. Rep.\/} {\bf
  474}, 1 (2009).

\bibitem{tothapellaniz14}
G.~T{\'o}th, I.~Apellaniz, Quantum metrology from a quantum information science
  perspective, {\it J. Phys. A: Math. Theo.\/} {\bf 47}, 424006 (2014).

\bibitem{2016arXiv160901609P}
L.~Pezz{\'e}, A.~Smerzi, M.~K. Oberthaler, R.~Schmied, P.~Treutlein,
  Non-classical states of atomic ensembles: fundamentals and applications in
  quantum metrology, {\it ArXiv e-prints: 1609.01609\/}  (2016).

\bibitem{Kitagawa1993Squeezed}
M.~Kitagawa, M.~Ueda, Squeezed spin states, {\it Phys. Rev. A\/} {\bf 47}, 5138
  (1993).

\bibitem{Wineland1994Squeezed}
D.~J. Wineland, J.~J. Bollinger, W.~M. Itano, D.~J. Heinzen, Squeezed atomic
  states and projection noise in spectroscopy, {\it Phys. Rev. A\/} {\bf 50},
  67 (1994).

\bibitem{Hald1999Spin}
J.~Hald, J.~L. S{\o}rensen, C.~Schori, E.~S. Polzik, Spin squeezed atoms: A
  macroscopic entangled ensemble created by light, {\it Phys. Rev. Lett.\/}
  {\bf 83}, 1319 (1999).

\bibitem{Fernholz2008Spin}
T.~Fernholz, H.~Krauter, K.~Jensen, J.~F. Sherson, A.~S. S{\o}rensen, E.~S.
  Polzik, Spin squeezing of atomic ensembles via nuclear-electronic spin
  entanglement, {\it Phys. Rev. Lett.\/} {\bf 101}, 073601 (2008).

\bibitem{Orzel2386}
C.~Orzel, A.~K. Tuchman, M.~L. Fenselau, M.~Yasuda, M.~A. Kasevich, Squeezed
  states in a bose-einstein condensate, {\it Science\/} {\bf 291}, 2386 (2001).

\bibitem{Riedel2010Atom-chip-based}
M.~F. Riedel, P.~B{\"o}hi, Y.~Li, T.~W. H{\"a}nsch, A.~Sinatra, P.~Treutlein,
  Atom-chip-based generation of entanglement for quantum metrology, {\it
  Nature\/} {\bf 464}, 1170 (2010).

\bibitem{Esteve2008Squeezing}
J.~Esteve, C.~Gross, A.~Weller, S.~Giovanazzi, M.~Oberthaler, Squeezing and
  entanglement in a bose--einstein condensate, {\it Nature\/} {\bf 455}, 1216
  (2008).

\bibitem{SchleierSmithPRL2010}
M.~H. Schleier-Smith, I.~D. Leroux, V.~{Vuleti\ifmmode \acute{c}\else
  {\'c}\fi{}}, States of an ensemble of two-level atoms with reduced quantum
  uncertainty, {\it Phys. Rev. Lett.\/} {\bf 104}, 073604 (2010).

\bibitem{Leroux2010}
I.~D. Leroux, M.~H. Schleier-Smith, V.~Vuleti{\'c}, Implementation of cavity
  squeezing of a collective atomic spin, {\it Phys. Rev. Lett.\/} {\bf 104},
  73602 (2010).

\bibitem{Gross2010Nonlinear}
C.~Gross, T.~Zibold, E.~Nicklas, J.~Esteve, M.~K. Oberthaler, Nonlinear atom
  interferometer surpasses classical precision limit, {\it Nature\/} {\bf 464},
  1165 (2010).

\bibitem{BohnetJ2014}
J.~G. Bohnet, K.~C. Cox, M.~A. Norcia, J.~M. Weiner, Z.~Chen, J.~K. Thompson,
  Reduced spin measurement back-action for a phase sensitivity ten times beyond
  the standard quantum limit, {\it Nat. Photon\/} {\bf 8}, 731 (2014).

\bibitem{SewellPRL2012}
R.~J. Sewell, M.~Koschorreck, M.~Napolitano, B.~Dubost, N.~Behbood, M.~W.
  Mitchell, Magnetic sensitivity beyond the projection noise limit by spin
  squeezing, {\it Phys. Rev. Lett.\/} {\bf 109}, 253605 (2012).

\bibitem{Cox2016}
K.~C. Cox, G.~P. Greve, J.~M. Weiner, J.~K. Thompson, Deterministic squeezed
  states with collective measurements and feedback, {\it Phys. Rev. Lett.\/}
  {\bf 116}, 093602 (2016).

\bibitem{HostenN2016}
O.~Hosten, N.~J. Engelsen, R.~Krishnakumar, M.~A. Kasevich, Measurement noise
  100 times lower than the quantum-projection limit using entangled atoms, {\it
  Nature\/} {\bf 529}, 505 (2016).

\bibitem{Meyer2001}
V.~Meyer, M.~A. Rowe, D.~Kielpinski, C.~A. Sackett, W.~M. Itano, C.~Monroe,
  D.~J. Wineland, Experimental demonstration of entanglement-enhanced rotation
  angle estimation using trapped ions, {\it Phys. Rev. Lett.\/} {\bf 86}, 5870
  (2001).

\bibitem{Auccaise2015}
R.~Auccaise, A.~G. Araujo-Ferreira, R.~S. Sarthour, I.~S. Oliveira, T.~J.
  Bonagamba, I.~Roditi, Spin squeezing in a quadrupolar nuclei nmr system, {\it
  Phys. Rev. Lett.\/} {\bf 114}, 043604 (2015).

\bibitem{MitchellExtremespinsqueezing2014}
M.~W. Mitchell, F.~A. Beduini, Extreme spin squeezing for photons, {\it New J.
  Phys.\/} {\bf 16}, 073027 (2014).

\bibitem{Sorensen2001Many-particle}
A.~S{\o}rensen, L.-M. Duan, J.~Cirac, P.~Zoller, Many-particle entanglement
  with bose--einstein condensates, {\it Nature\/} {\bf 409}, 63 (2001).

\bibitem{Sorensen2001Entanglement}
A.~S. S{\o}rensen, K.~M{\o}lmer, Entanglement and extreme spin squeezing, {\it
  Phys. Rev. Lett.\/} {\bf 86}, 4431 (2001).

\bibitem{Lucke2014Detecting}
B.~L{\"u}cke, J.~Peise, G.~Vitagliano, J.~Arlt, L.~Santos, G.~T{\'o}th,
  C.~Klempt, Detecting multiparticle entanglement of dicke states, {\it Phys.
  Rev. Lett.\/} {\bf 112}, 155304 (2014).

\bibitem{vitagliano16}
G.~Vitagliano, I.~Apellaniz, M.~Kleinmann, B.~L{\"u}cke, C.~Klempt, G.~Toth,
  Entanglement and extreme spin squeezing of unpolarized states, {\it New J.
  Phys.\/} {\bf 19} (2017).

\bibitem{Wieczorek2009Experimental}
W.~Wieczorek, R.~Krischek, N.~Kiesel, P.~Michelberger, G.~T{\'o}th,
  H.~Weinfurter, Experimental entanglement of a six-photon symmetric dicke
  state, {\it Phys. Rev. Lett.\/} {\bf 103}, 020504 (2009).

\bibitem{Prevedel2009Experimental}
R.~Prevedel, G.~Cronenberg, M.~S. Tame, M.~Paternostro, P.~Walther, M.~S. Kim,
  A.~Zeilinger, Experimental realization of dicke states of up to six qubits
  for multiparty quantum networking, {\it Phys. Rev. Lett.\/} {\bf 103}, 020503
  (2009).

\bibitem{Lucke2011Twin}
B.~L{\"u}cke, M.~Scherer, J.~Kruse, L.~Pezz{\'e}, F.~Deuretzbacher, P.~Hyllus,
  J.~Peise, W.~Ertmer, J.~Arlt, L.~Santos, A.~Smerzi, C.~Klempt, Twin matter
  waves for interferometry beyond the classical limit, {\it Science\/} {\bf
  334}, 773 (2011).

\bibitem{Hamley2012Spin-nematic}
C.~Hamley, C.~Gerving, T.~Hoang, E.~Bookjans, M.~Chapman, Spin-nematic squeezed
  vacuum in a quantum gas, {\it Nat. Phys.\/} {\bf 8}, 305 (2012).

\bibitem{Hoang2016Characterizing}
T.~M. {Hoang}, M.~{Anquez}, M.~J. {Boguslawski}, H.~M. {Bharath}, B.~A.
  {Robbins}, M.~S. {Chapman}, Characterizing the energy gap and demonstrating
  an adiabatic quench in an interacting spin system, {\it arXiv:1512.06766\/}
  (2015).

\bibitem{Luo620}
X.-Y. Luo, Y.-Q. Zou, L.-N. Wu, Q.~Liu, M.-F. Han, M.~K. Tey, L.~You,
  Deterministic entanglement generation from driving through quantum phase
  transitions, {\it Science\/} {\bf 355}, 620 (2017).

\bibitem{HePRA2011}
Q.~Y. He, S.-G. Peng, P.~D. Drummond, M.~D. Reid, Planar quantum squeezing and
  atom interferometry, {\it Phys. Rev. A\/} {\bf 84}, 022107 (2011).

\bibitem{He2012}
Q.~Y. He, T.~G. Vaughan, P.~D. Drummond, M.~D. Reid, Entanglement, number
  fluctuations and optimized interferometric phase measurement, {\it New J.
  Phys.\/} {\bf 14}, 093012 (2012).

\bibitem{PuentesNJP2013}
G.~Puentes, G.~Colangelo, R.~J. Sewell, M.~W. Mitchell, Planar squeezing by
  quantum non-demolition measurement in cold atomic ensembles, {\it New J.
  Phys.\/} {\bf 15}, 103031 (2013).

\bibitem{ColangeloNat2017}
G.~Colangelo, F.~M. Ciurana, L.~C. Bianchet, R.~J. Sewell, M.~W. Mitchell,
  Simultaneous tracking of spin angle and amplitude beyond classical limits,
  {\it Nature\/} {\bf 543}, 525 (2017).

\bibitem{ColangeloPQS2017}
G.~Colangelo, F.~{Martin Ciurana}, G.~Puentes, M.~W. Mitchell, R.~J. Sewell,
  Entanglement-enhanced phase estimation without prior phase information, {\it
  Phys. Rev. Lett.\/} {\bf 118}, 233603 (2017).

\bibitem{Guhne2005Multipartite}
O.~G{\"u}hne, G.~T{\'o}th, H.~J. Briegel, Multipartite entanglement in spin
  chains, {\it New J. Phys.\/} {\bf 7}, 229 (2005).

\bibitem{Note1}
\label {footnote_supplement}See Supplemental Material at [URL].

\bibitem{Note2}
Note that the S\o rensen-M\o lmer condition requires as input just one
  variance, that must be the one orthogonal to the polarization.

\bibitem{PhysRevLett.98.110502}
O.~G{\"u}hne, M.~Reimpell, R.~F. Werner, Estimating entanglement measures in
  experiments, {\it Phys. Rev. Lett.\/} {\bf 98}, 110502 (2007).

\bibitem{1367-2630-9-3-046}
J.~Eisert, F.~G. S.~L. Brand{\~a}o, K.~M.~R. Audenaert, Quantitative
  entanglement witnesses, {\it New Journal of Physics\/} {\bf 9}, 46 (2007).

\bibitem{DammeierNJP2015}
L.~Dammeier, R.~Schwonnek, R.~F. Werner, Uncertainty relations for angular
  momentum, {\it New J. Phys.\/} {\bf 17}, 093046 (2015).

\bibitem{2017arXiv170806986M}
O.~{Marty}, M.~{Cramer}, G.~{Vitagliano}, G.~{Toth}, M.~B. {Plenio},
  Multiparticle entanglement criteria for nonsymmetric collective variances,
  {\it ArXiv e-prints\/}  (2017).

\bibitem{Colangelo2013a}
G.~Colangelo, R.~J. Sewell, N.~Behbood, F.~M. Ciurana, G.~Triginer, M.~W.
  Mitchell, Quantum atom--light interfaces in the gaussian description for
  spin-1 systems, {\it New J. Phys.\/} {\bf 15}, 103007 (2013).

\end{thebibliography}

\clearpage 

\renewcommand{\thefigure}{S\arabic{figure}}
\renewcommand{\thetable}{S\arabic{table}}
\renewcommand{\theequation}{S\arabic{equation}}
\setcounter{figure}{0}
\setcounter{table}{0}
\setcounter{equation}{0}
\setcounter{page}{1}
\thispagestyle{empty}

\onecolumngrid
\begin{center}
{\large \bf Supplemental Material for \\``Entanglement and extreme planar spin squeezing''}

\bigskip
G.~Vitagliano$^{1,2},$ G.~Colangelo$^{3},$ F.~Martin Ciurana$^{3},$ M.~W.~Mitchell$^{3,4},$ R.~J. Sewell$^{3},$ G.~T\'oth$^{2,5,6}$

\smallskip
{\it \small
$^1$Institute for Quantum Optics and Quantum Information (IQOQI),\\ Austrian Academy of Sciences, Boltzmanngasse 3, A-1090 Vienna, Austria

$^2$\BilbaoAddress

$^3$ICFO-Institut de Ciencies Fotoniques, The Barcelona Institute of \\ Science and Technology, 08860 Castelldefels (Barcelona), Spain

$^4$\ICREAAddress

$^5$\IKERBASQUEAddress

$^6$\HungaryAddress
}

(Dated: \today)

\medskip
\medskip

\parbox[b][1cm][t]{0.85\textwidth}{\quad
The supplemental material contains some details of the calculation obtaining convex hulls, and of the proof of Observation 3. We also present a brief description of the experiment creating planar squeezed states.
}

\medskip
\medskip
\medskip

\end{center}

\twocolumngrid

{\it Convex hull $\mathcal G_k^{(j)}$ from spin-$J$ subspaces.---}The problem of finding the convex hull $\mathcal G_k^{(j)}$ of the optimal function defined by Eq.~(\ref{eq:hkjfunc}) can be approached exploiting the Legendre transform, in this framework defined as \cite{PhysRevLett.98.110502,1367-2630-9-3-046}
\begin{equation}\label{eq:legendre1}
\mathcal L[(\Delta L_\parallel)_\phi^2/kj](T) := \inf_\phi [\tfrac 1 {kj}(\Delta L_\parallel)_\phi^2-\ave{T}_\phi ] ,
\end{equation}
for the normalized planar variance $(\Delta L_\parallel)_\phi^2/kj$
as a function of a certain observable $T$, which we will choose as $T=L_y/kj$. Then, based the definition \eqref{eq:legendre1}, we can find the lower bound $(\Delta L_\parallel)_\phi^2 \geq \mathcal G_k^{(j)}(X)$ by means of another Legendre transform 
\begin{equation}\label{eq:convhullfunct}
\mathcal G_k^{(j)}(X) := \sup_\lambda \{\lambda X - \mathcal L[(\Delta L_\parallel)_\phi^2/kj](\lambda L_y/kj)\} ,
\end{equation}
where $X$ is a real number. The bound \eqref{eq:convhullfunct}
is precisely the convex hull that we are looking for. Furthermore, we can also write the first Legendre transform \eqref{eq:legendre1} as an eigenvalue problem (see also \cite{DammeierNJP2015,2017arXiv170806986M} for a general addessing of similar problems)
\begin{equation}
\label{eq:legendretransfeig}
\mathcal L[(\Delta L_\parallel)_\phi^2/kj](\lambda L_y/kj) 
=\tfrac 1 {kj} \min_{s_y,s_z}[ \min_\phi \ave{H_{s_y,s_z,\lambda}}_\phi ],
\end{equation}
where the parametric Hamiltonian $H_{s_y,s_z,\lambda}=(L_y-s_y)^2+(L_z-s_z)^2-\lambda L_y$ is a collective operator acting on a $k$-partite space of spin $j$ particles. Let us write a general pure state as $\ket{\phi}=\sum_{J=0}^{kj} a_J \ket{\psi_J},$ which is a superposition of single spin-$J$ states $\ket{\psi_J}$. Since the collective operator can be decomposed as a direct sum of single spin $J$ operators with $0\leq J \leq kj$ ($1/2\leq J \leq kj$ for half-integer $j$ and odd $k$) the optimization can be simplified.  The expectation value in Eq.~(\ref{eq:legendretransfeig}) can be written as the following sum
\begin{equation}\label{eq:HintermsofJ}
\ave{H_{s_y,s_z,\lambda}}_\phi = \sum_{J=0}^{kj} a_J^2 \ave{(L^{(J)}_y-s_y)^2+(L^{(J)}_z-s_z)^2-\lambda L^{(J)}_y}_{\psi_J},
\end{equation}
where the $L^{(J)}_m$ are single spin-$J$ operators. Let us restrict ourselves for simplicity to $k>1$ and the cases of integer $j$ or half-integer $j$ and even $k$. In such cases the space of $k$ particle states includes the singlet (i.e., a $J=0$ state) and we can easily prove the property $\mathcal G_k^{(j)}(0)=0$. In fact, the value $\mathcal G_k^{(j)}(0)=0$ can be reached by choosing $\ket{\phi}$ in Eq.~(\ref{eq:convhullfunct}) as the singlet, which is also the infimum, since the function $\mathcal G_k^{(j)}(X)$ must be positive for all $X$.

More in general, substituting Eqs.~(\ref{eq:legendretransfeig}) and (\ref{eq:HintermsofJ}) into Eq.~(\ref{eq:convhullfunct}) one can see that the function $\mathcal G_k^{(j)}(X)$ can be
obtained with minimizations in the single spin-$J$ subspaces with $0\leq J\leq kj$. Thus, clearly by increasing $k$ one has to minimize over a larger number of subspaces and consider a higher  number of parameters $a_J$, which makes the resulting function decreasing wiht $k$. On the other hand, for computing just $\mathcal G_{\rm sy}^{(J)}(X)$ for a certain $J$ one has to put $a_J=1$ and thus restrict the optimization to a single spin-$J$ subspace.

{\it Proof of Observation 3: Quantifying the fraction of particles in $(k+1)$-entangled groups.---}Let us consider the entanglement depth criterion written in the form (\ref{eq:plansqcritkprovv}) as in Observation 1.
Given a certain $k$, we can give a quantitative interpretation of the degree of violation of the criterion by providing an estimate of the minimal fraction of particles in $(k+1)$-entangled groups. 

Let us consider a pure state of $N$ particles $|\Phi_{\mathcal N}\rangle = \bigotimes_{n=1}^{\mathcal N} |\phi_n\rangle \otimes |\Psi_{\rm rest}\rangle$ for some partition that contains $\mathcal N$ groups of $k_n \leq k$ particles, with $\sum_{n=1}^{\mathcal N} k_n = M_{\mathcal N}$ and the rest in a collective state $|\Psi_{\rm rest}\rangle$ of $N-M_{\mathcal N}$ particles that are entangled in groups of $k+1$ or more. 
For such a state we have $(\Delta J_\parallel)_{\mathcal N}^2 \geq \sum_{n=1}^{\mathcal N} (\Delta L_\parallel)_{\phi_n} \geq  M_{\mathcal N}j \mathcal G_k^{(j)}\left(\sum_{n=1}^{\mathcal N} \ave{L_y}_{\phi_n}/M_{\mathcal N}j \right)$, due to convexity and the fact that $\mathcal G_k^{(j)}(X)\geq \mathcal G_{k_n}^{(j)}(X)$ for $k_n\leq k$.

At this point, we assume that $\langle L_y\rangle$ is distributed among the 
 $\mathcal N$ groups and the rest of the particles in proportion of the number of particles in these two groups, i.e., 
$\sum_{n=1}^{\mathcal N} \ave{L_y}_{\phi_n}/M_{\mathcal N}j = \ave{J_y}_{\mathcal N}/Nj.$
Hence, we arrive at $(\Delta J_\parallel)_{\mathcal N}^2 \geq M_{\mathcal N}j \mathcal G_k^{(j)}\left( \ave{J_y}_{\mathcal N}/Nj \right)$.
Due to the concavity of the variance and the convexity of $\mathcal G_k^{(j)}(X)$ this inequality also holds for mixtures of 
states of the type $|\Phi_{\mathcal {N}}\rangle$ with a fixed particle number $N,$ denoted by $\varrho_N.$ Hence, we obtain $(\Delta J_\parallel)_{\varrho_{N}}^2 \geq \ave{M}_{\varrho_{ N}}j \mathcal G_k^{(j)}\left( \ave{J_y}_{\varrho_{ N}}/Nj \right).$

Now we consider states $\varrho=\sum_{ N} r_{ N} \varrho_{ N}$, where $r_{ N}$ are probabilities associated with different number of particles $N$ and groupings, and define $Q=\ave{M}_\varrho/\ave{N}_\varrho$, where  $\ave{N}_\varrho=\sum_{ N} r_{N} N$ is the average particle number, and $\ave{M}_\varrho=\sum_{ N}  r_{N} \ave{M}_{ N}.$
We have $(\Delta J_\parallel)_{\varrho}^2 \geq \sum_{N} r_{ N} (\Delta J_\parallel)_{N}^2 \geq \sum_{ N} r_{ N} \ave{M}_{N}j \mathcal G_k^{(j)}\left( \ave{J_y}_{N}/Nj \right)=Q\sum_{N} r_{N} Nj \mathcal G_k^{(j)}\left( \ave{J_y}_{N}/Nj \right)$ and by using Jensen inequality we arrive at $(\Delta J_\parallel)_{\varrho}^2 \geq Q\ave{N}_\varrho j \mathcal G_k^{(j)}\left( \ave{J_y}_{\varrho} /\ave{N}_\varrho j \right)$.
Using Eqs.~\eqref{eq:xi_parallel} and \eqref{eq:lowboundJ}, Observation 3 follows. 
$\qed$

So far, in the derivations we made the assumption that the total polarization splits equally for the different sub-ensembles of atoms.
Without such an assumption, first for pure states, we analyze the worst-case scenario in which for a state like $|\Phi_{\mathcal N}\rangle$ the polarization splits unequally and the state $|\Psi_{\rm rest}\rangle$ is polarized as much as possible. Hence, we assume $\ave{J_y}_{\Psi_{\rm rest}}=(N-M_{\mathcal N})j$ and it follows that
$\sum_{n=1}^{\mathcal N} \ave{L_y}_{\phi_n} = \ave{J_y}_{\mathcal N}-(N-M_{\mathcal N})j$, and consequently
$(\Delta J_\parallel)_{{\mathcal N}}^2 \geq M_{\mathcal N}j \mathcal G_k^{(j)}\left[(\ave{J_y}_{\mathcal N}-(N-M_{\mathcal N})j)/M_{\mathcal N}j\right].$
Using Eq.~\eqref{eq:lowboundJ}, we obtain 
$(\Delta J_\parallel)_{{\mathcal N}}^2\ge \zeta^2_J  [\ave{J_y}_{\mathcal N}-(N-M_{\mathcal N})j].$
This is clearly valid for mixed states with a varying particle number as 
$(\Delta J_\parallel)_{\varrho}^2\ge \zeta^2_J  [\ave{J_y}_{\varrho}-(\ave{N}_{\varrho}-\ave{M}_{\varrho})j],$
which can further be rewritten as 
$Q \le (\xi^2_{\parallel}/\zeta^2_J+W-1)/W$ where $W=\ave{N}_{\varrho}j/\ave{J_y}_{\varrho}$. 
Then, for a state that is almost fully polarized, i.e., 
$\ave{J_y}_{{\mathcal N}}\approx \ave{N}_{\mathcal N} j$  we recover the statement of Observation 3.

Argument similar to the proof of Observation~3 above can be applied also to 
the S\o rensen-M\o lmer criterion, which states that 
\begin{equation}(\Delta J_z)^2 \ge Nj F_{J}\left(\frac{\ave{J_y}}{Nj}\right) \end{equation} holds in a system of spin-$j$ particles for states with an entanglement depth of at most  $J/j.$ Here, $F_J(X)$ is a convex function obtainend numerically ~\cite{Sorensen2001Entanglement}. 
Based on our discussion, we can interpret the degree of violation of the S\o rensen-M\o lmer criterion in a similar way.

\begin{figure}[t!]
\includegraphics[width=\columnwidth ]{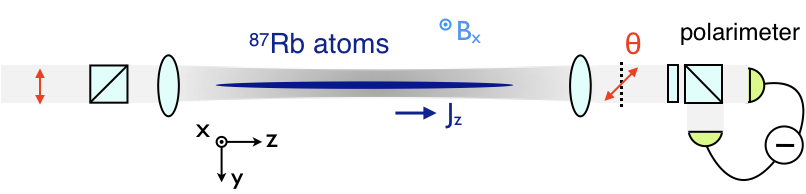}
\caption{The experimental configuration of Ref.~\cite{ColangeloPQS2017}: an ensemble of laser-cooled $^{87}$Rb atoms trapped in a singe-beam optical dipole trap and precessing in the $y$--$z$ plane due to an external magnetic field $B_x$.
The atoms are probed via paramagnetic Faraday rotation: the polarization of input linearly polarized optical pulses rotates by an angle $\theta \propto J_z$, the spin projection onto the measurement axis, as it passes through the atoms, and is detected using a balanced polarimeter.
\label{fig:state} }
\end{figure}

{\it Description of Experiment.---}Experimental data is taken from Ref.~\cite{ColangeloPQS2017}.
In this experiment, an ensemble of laser-cooled, spin-$1$ Rb$^{87}$ atoms was loaded into a single-beam optical dipole trap, polarized via optical pumping, and allowed to precess in the $(y,z)$-plane under an external magnetic field at a rate $\omega_{\rm L}\simeq2\pi\times26$~kHz.
Measurement-induced spin squeezing of the $(\Delta J_y)^2$ and $(\Delta J_z)^2$ was achieved via Faraday rotation probing using a train of near-resonant, $\mu$s-duration optical pulses.

More concretely, the collective spin oscillates such that $J_z(t) = J_z \cos \phi - J_y \sin \phi $, where $\phi=\omega_{\rm L} t$.
The atoms and light interact via the hamiltonian $H = g S_z J_z(t)$, where the Stokes operators $S_k$ describe the optical polarization.
This describes a quantum non-demolition (QND) measurement of instantaneous spin projection $J_z(t)$: the optical polarization rotates by an angle $\theta = g J_z(t)$, where $g$ is a coupling constant, proportional to the instantaneous spin projection along the $z$-axis~\cite{Colangelo2013a}.
Measurement of $\theta$ projects the atoms onto a state with $(\Delta J_z(t))^2$ reduced by a factor $\sim 1/(1+g^2 N n)$, where $N$ is the number of atoms in the ensemble, and $n$ is the number of photons in a single probe pulse.
Correspondingly, $(\Delta J_x(t))^2$ is increased by a factor $\sim 1 + g^2 n$, and $(\Delta J_y(t))^2$ is increased by a negligible factor of order $1$.
Repeated QND measurements of $J_z(t)$ as the spins oscillate progressively squeezes the input $J_z$ and $J_y$ spin components, to produce the PQS state.
At the same time, off-resonant scattering of probe photons during the measurement leads to decay of the spin polarization at a rate $\eta \propto g^2$, and introduces noise $\beta \propto n$ into the atomic spin components.
This leads to a trade-off between measurement-induced squeezing and decoherence, and an optimum measurement strength, characterized by the total photon number $N_{\rm L} = p n$, where $p$ is the number of probe pulses.

In the experiment, the PQS state was detected by recording a series of measurements $\theta(t_k)$ and fitting using a free induction decay model
\begin{equation}
\label{eq:FIDForm}
\theta(t)=g \Big[ J_z(\tmeas) \cos \phi - J_y(\tmeas) \sin \phi \Big] e^{-\trel/T_2} + \theta_0,
\end{equation}
where $\trel \equiv t- \tmeas$ and the phase $\phi=\omegaL \trel$.
This model allows a simultaneous estimation of $\J = \{J_z(\tmeas),J_y(\tmeas)\}$ producing a conditional PQS state at time $\tmeas$.
$\tmeas$ can be adjusted, allowing to study how the spin squeezing and entanglement evolves during the measurement as a function of $N_{\rm L}$.
Conditional spin squeezing was detected by comparing two estimates, $\J_1$ and $\J_2$, taken from the set of measurements immediately before and after $\tmeas$, and computing the conditional covariance matrix  $\Gamma_{\J_2\mid \J_1}=\Gamma_{\J_2}-\Gamma_{\J_{2} \J_1} \Gamma_{\J_1}^{-1}\Gamma_{\J_1 \J_2}$ which quantifies the error in the best linear prediction of $\J_2$ based on $\J_1$, where $\Gamma_{\bf v}$ indicates the covariance matrix for vector ${\bf v}$, and $\Gamma_{\bf uv}$ indicates the cross-covariance matrix for vectors ${\bf u}$ and ${\bf v}$.
The measurement sequence was repeated 453 times to acquire statistics.
Measurement read-out noise $\Gamma_0$ was quantified by repeating the measurement sequence without atoms in the trap.
The atomic spin covariance matrix was then estimated as $\Gamma = \Gamma_{\J_2\mid \J_1}-\Gamma_0$, which has entries $ \Gamma_{ij} = \langle J_iJ_j +J_jJ_i\rangle/2 - \langle J_i\rangle \langle J_j\rangle$.

\end{document}